\documentclass[12pt,reqno]{article}

\usepackage{amssymb}
\usepackage{amsmath}
\usepackage{amsthm}
\usepackage{bm}
\usepackage{bbm}
\usepackage {mathtools}

\usepackage{xargs}
\usepackage{xparse}

\usepackage[colorlinks=true,linkcolor=teal,urlcolor=teal,citecolor=teal]{hyperref}
\newcommand{\refcs}[2]{\hyperref[#2]{\ref*{#1}.\ref*{#2}}}
\newcommand{\eqrefcs}[2]{\hyperref[#2]{(\ref*{#1}.\ref*{#2})}}

\usepackage{subfig}
\usepackage{graphicx}

\usepackage{a4wide}
\usepackage[bottom=1in,top=1in]{geometry}
\usepackage{authblk}

\usepackage{color,xcolor}

\usepackage[utf8]{inputenc}
\usepackage[T1]{fontenc}

\usepackage{subdepth}
\usepackage{stackengine}

\usepackage{enumitem}
\setlist[itemize]{label=\textbullet}

\usepackage[prependcaption]{todonotes}
\newcommandx{\enligne}[2][1=]{\todo[inline,linecolor=yellow,backgroundcolor=yellow!25,bordercolor=yellow,#1]{#2}}
\newcommandx{\aubord}[2][1=]{\todo[linecolor=yellow,backgroundcolor=yellow!25,bordercolor=yellow,#1]{#2}}

\newcommand{\cc}{\mathbb{C}}
\newcommand{\rr}{\mathbb{R}}

\newcommand{\cB}{\mathcal{B}}

\renewcommand{\d}{\mathrm{d}}
\renewcommand{\i}{\mathrm{i}}

\newcommand{\ind}{\mathbbm{1}}
\newcommand{\id}{\mathrm{Id}}

\DeclareMathOperator{\ran}{Ran}

\DeclareMathOperator{\tr}{tr}

\newcommand{\bra}[1]{\langle#1|}
\newcommand{\ket}[1]{|#1\rangle}
\newcommand{\ketbra}[2]{\ket{#1}\!\bra{#2}}
\newcommand{\braket}[2]{ \langle #1 , #2 \rangle}

\DeclareDocumentCommand \aT { o } {%
  \IfNoValueTF {#1} {%
    \alpha(\test)%
  }{%
    \alpha(\test_{#1})%
  }%
}
\DeclareDocumentCommand \bT { o } {%
  \IfNoValueTF {#1} {%
    \beta(\test)%
  }{%
    \beta(\test_{#1})%
  }%
}
\newcommand{\drs}{\Delta_{\rho|\sigma}}
\newcommand{\orho}{\Omega_{\rho}}
\newcommand{\osig}{\Omega_{\sigma}}
\newcommand{\oome}{\Omega_{\omega}}
\newcommand{\pp}{\mathfrak{P}}
\newcommand{\test}{T}
\newcommand{\M}{\mathcal{M}}
\newcommand{\N}{\mathcal{N}}

\renewcommand{\H}{\mathcal{H}}
\newcommand{\K}{\mathcal{K}}
\newcommand{\norme}[1]{{\|#1\|}}
\newcommand{\spann}{\mathop{\rm span}\nolimits}
\newcommand{\Lp}[1]{\mathrm{L}^{#1}}
\newcommand{\bh}{\cB(\H)}
\newcommand{\J}{\mathrm{J}}
\renewcommand{\L}{\mathrm{L}}
\newcommand{\erho}{\mathrm{e}_{\rho}}
\newcommand{\esig}{\mathrm{e}_{\sigma}}
\newcommand{\tesig}{\tilde{\mathrm{e}}_\sigma}
\newcommand{\proj}{P}
\newcommand{\sproj}{p}
\newcommand{\np}{_{\mathrm{NP}}}
\newcommand{\kl}{_{\mathrm{KL}}}

\renewcommand{\tfrac}{\genfrac{}{}{}1}

\newtheoremstyle{perso}
  {\topsep}   
  {\topsep}   
  {\itshape}  
  {0pt}       
  {\scshape} 
  {.}         
  {5pt plus 1pt minus 1pt} 
  {}          

\theoremstyle{perso}
	\newtheorem{defi}{Definition}
	\newtheorem{theo}[defi]{Theorem}
	\newtheorem{lemm}[defi]{Lemma}

\theoremstyle{remark}
	\newtheorem{rema}[defi]{Remark}

\newenvironment{myitemize}
	{\begin{itemize}
	    \setlength{\itemsep}{1pt}
	    \setlength{\parskip}{1pt}
	    \setlength{\parsep}{1pt}}{\end{itemize}}

\usepackage[nocompress]{cite}

\usepackage{tikz}
	\usetikzlibrary{positioning}
	\usetikzlibrary{math}
	\usetikzlibrary{patterns}

\begin{document}
\title{Ke Li's lemma for quantum hypothesis testing in general von Neumann algebras
}
\author{Yan Pautrat}
\affil{Université Paris-Saclay, CNRS, Laboratoire de Mathématiques d'Orsay,
	91405 Orsay, France}
\author{Simeng Wang}
\affil{Institute for Advanced Study in Mathematics, Harbin Institute of Technology, Harbin 150001, China.}
\maketitle

\begin{abstract}
A lemma stated by Ke Li in \cite{KeLi} has been used in e.g.\cite{DPR,DR,KW17,WTB,TT15} for various tasks in quantum hypothesis testing, data compression with quantum side information or quantum key distribution. This lemma was originally proven in finite dimension, with a direct extension to type I von Neumann algebras. Here we show that the use of modular theory allows to give more transparent meaning to the objects constructed by the lemma, and to prove it for general von Neumann algebras. This yields a new proof of quantum Stein's lemma with slightly weaker assumption, as well as immediate generalizations of its second order asymptotics, for example the main results in \cite{DPR} and \cite{KeLi}. 
\end{abstract}

\section{Introduction} 
\label{sec:introduction}

Quantum hypothesis testing is concerned with the situation where one considers a von Neumann algebra $\M$, 
equipped with a state which is either $\rho$ or $\sigma$; this uncertainty in the nature of the state makes sense in particular when $\M$ is viewed as modeling the observable quantities of a quantum system, and the physical state of this system --- itself modeled by a state in the mathematical sense --- is only known to be one or the other. A natural task is then to try and determine which state is the true one by producing, in physical terms, an experimental measurement procedure such that, depending on the measurement outcome, one will conclude that the actual state is $\rho$ or $\sigma$.

In the model of orthodox quantum mechanics, a measurement procedure is determined by a self-adjoint element $X$ of the observable algebra $\M$, and this $X$ is simply called an \emph{observable}. Non-trivial measurements will have a random outcome; the set of possible outcomes is exactly the spectrum of that element, and if $\omega$ is the actual state of the system and we denote by $\xi_X$ the spectral measure of $X$, then the probability distribution for the measurement outcomes is $\omega\circ\xi_X$. In the situation described above it will suffice to consider an observable with spectrum $\{0,1\}$, that is, an orthogonal projector of $\M$. We will take advantage of this simplification and continue this preliminary discussion assuming that the observable describing the discriminating experience, is an orthogonal projector $\test$, which we call a \emph{test}.

Suppose the decision rule is that, if the measurement outcome is $0$, then the observer concludes that the true state is $\rho$, and if the measurement outcome is $1$, then they conclude that the true state is $\sigma$. There are two ways in which this conclusion can be wrong: either the actual state was $\rho$ and the measurement outcome was $1$, or the actual state was $\sigma$ and the measurement outcome was $0$. Applying the rules determining the distribution of the measurement outcomes shows that the former occurs with probability $\rho(\test)$, and the latter with probability $\sigma(\id-\test)$. Assuming that the possible states $\rho$ and $\sigma$ are fixed, we denote these two types or error
\begin{equation} \label{eq_defalphabeta}
  \aT:=\rho(\test)\qquad \bT:=\sigma(\id-\test).
\end{equation}
If, to the experimenter's knowledge, both $\rho$ and $\sigma$ may be the state of the system, then one typically wishes to make both $\aT$ and $\bT$ small. There are, however, various ways in which this can be done, and we postpone the corresponding discussion to section~\ref{sec:discussion}.

The result we are concerned with gives a test $T$ such that $\aT$ and $\bT$ satisfy a pair of upper bounds. To state it, let us assume that $\M$ is a finite-dimensional matrix algebra: $\M=\cB(\cc^n)$, and denote by $\varrho$ and $\varsigma$ the density matrices associated with the states $\rho$ and $\sigma$, that is:
\begin{equation}\label{eq_matdens}
  \rho = \tr(\varrho\,\cdot) \qquad \sigma= \tr(\varsigma\,\cdot).
\end{equation}
Assume for simplicity that $\rho$ and $\sigma$ are faithful states, or equivalently that $\varrho$, $\varsigma$ are invertible matrices. Consider then the vector space $\cB(\cc^n)$, on which we define the vectors 
\[\orho = \varrho^{1/2}\qquad \osig=\varsigma^{1/2}\]
and the operator
\[\drs : X \mapsto \varrho X \varsigma^{-1}.\]
If $\cB(\cc^n)$ is equipped with the scalar product $\braket XY = \tr(X^*Y)$ then $\drs$ is self-adjoint. It then holds that for any $\epsilon>0$ there exists a test $\test$ such that
\begin{equation}\label{eq_keli}
  \aT\leq \epsilon\qquad \bT\leq \braket{\osig}{\ind_{(\epsilon,+\infty)}(\drs)\osig},
\end{equation}
where $\ind_{(\epsilon,+\infty)}(T )$ for an operator $T$ denotes the spectral projection of $T$ corresponding to the interval $(\epsilon , +\infty)$.

This result was first proven in \cite{KeLi}, as a technical step towards the ``second order Stein's lemma'' which we discus in section \ref{sec:discussion}. It is generally quoted as ``Ke Li's lemma'' for quantum hypothesis testing, even though it is not the only result of Ke Li relevant to this field. The statement as written above, however, does not appear in \cite{KeLi} and in particular, there is no mention of the operator $\drs$ in that article. The result was later reformulated in \cite{DPR} (with the same goal of proving the second order Stein's lemma) to involve the operator $\drs$, which the reader may already have recognized to be the finite-dimensional instance of a \emph{relative modular operator}.

The relevance of rewriting the statement of Ke Li's lemma in terms of $\drs$, as was done in \cite{DPR}, derives from a standard rule of thumb discussed in \cite{JOPP}: if the proof of a finite-dimensional statement can be rewritten solely in terms of modular theory on $\cB(\cc^n)$, then that proof should extend with minimal effort to general von Neumann algebras. A similar progress was carried out in \cite{JOPS} where a modular-theoretic proof of an inequality from \cite{Audenaert_et_al} led to a proof of the so-called Chernoff bound, Hoeffding bound and (first order) Stein's lemma in general von Neumann algebras (or, to be more precise, to the ``existence'' part of this lemma --- another inequality is required to prove optimality). Note, however, that \cite{DPR} only managed to rewrite the \emph{statement} of Ke Li's lemma in modular terms, and that the proof still used an explicit decomposition of the trace-class operators $\varrho$ and $\varsigma$ from expression \eqref{eq_matdens}, a step which the authors of \cite{DPR} couldn't translate to the more general case. For this reason, until now the only generalization of Ke Li's original result beyond the finite-dimensional case was the extension in \cite{Khabbazi_Oskouei_2019} to separable, type I, von Neumann algebras. The present paper shows an extension of Ke Li's lemma holds for any two normal states $\rho$, $\sigma$ on a general von Neumann algebra. 
\smallskip

To state our extension we need to recall how the relative modular operator $\drs$ is defined in the general case, and we therefore postpone its statement to Section~\ref{sec:a_general_proof}. For now, let us recall the definition of $\test$ in the proof of \eqref{eq_keli} as in \cite{KeLi,DPR} in order to underline its connection with modular theory, and motivate our definition of $T$ in the more general statement.
Consider spectral decompositions of $\varrho$, $\varsigma$:
\[  \varrho=\sum_{x=1}^n\lambda_x \ketbra{a_x}{a_x}\qquad 
\varsigma=\sum_{y=1}^n\mu_y \ketbra{b_y}{b_y}
\]
where $(\lambda_x)_x$ and $(\mu_y)_y$ are labeled in nondecreasing order, and both families $(a_x)_x$ and $(b_y)_y$ are orthonormal bases of $\cc^n$. Define for any $y=1,\ldots,n$:
\[Q_{y}=\ind_{[0,\epsilon \mu_y]}(\varrho)= \sum_{x=1}^n \ind_{\lambda_x \leq \epsilon\mu_y} \ketbra{a_x}{a_x},\]
so that $(Q_y)_y$ is a nondecreasing family of projectors, and
\[\xi_{y}=Q_{y}b_y = \sum_{x=1}^n \ind_{\lambda_x \leq \epsilon\mu_y} \braket{a_x}{b_y} a_x.\]
Ke Li then defines his test as the orthogonal projector $\test\kl$ onto the vector space spanned by the $\xi_y$, $y=1,\ldots,n$.

For $X$ an operator on $\cc^n$ denote now by $\L_X$ the operator on $\cB(\cc^n)$ acting by $Y\mapsto XY$, and let $\J$ be the antilinear involution $X\mapsto X^*$ on $\cB(\cc^n)$. Notice then that $(\J \L_X \J) Y = Y X^*$ and let $\M$, $\M'$ be the sets of operators of the form $\L_X$ or $\J \L_X\J$ respectively. We then have:
\[\Delta_{\rho|\sigma}=\sum_{x,y=1}^n {\lambda_x}{\mu_y}^{-1}\, \L_{\ketbra{a_x}{a_x}}\,\J \L_{\ketbra{b_y}{b_y}}\J\]
so that
\begin{equation*}\label{eq_2eexpp}
\ind_{(0,\epsilon]}(\Delta_{\rho|\sigma})=\sum_{x,y=1}^n \ind_{\lambda_x \leq \epsilon\mu_y}\, \L_{\ketbra{a_x}{a_x}}\,\J \L_{\ketbra{b_y}{b_y}}\J=\sum_{y=1}^n \L_{Q_y} \, \J \L_{\ketbra {b_y}{b_y}}\J
\end{equation*}
and
\begin{equation} \label{eq_pomegarho}
\ind_{(0,\epsilon]}(\Delta_{\rho|\sigma})\,\osig=\sum_{x,y=1}^n \ind_{\lambda_x\leq \epsilon \mu_y}\, \mu_y^{1/2} \braket{a_x}{b_y} \,{\ketbra{a_x}{b_y}}.
\end{equation}
It is then immediate to remark that $\ind_{(0,\epsilon]}(\Delta_{\rho|\sigma})\,\osig\, b_y=\mu_y^{1/2} \xi_y$ for any $y=1,\ldots,n$, so that
\begin{align*}
  \ran\big(\ind_{(0,\epsilon]}(\Delta_{\rho|\sigma})\,\osig\big)
  &=\spann\big\{\xi_y,\ y=1,\ldots,n\big\}
\end{align*}
where the last equation is due to the faithfulness of $\sigma$. Last, remark that the range $\ran\big(\ind_{(0,\epsilon]}(\Delta_{\rho|\sigma})\,\osig\big)$ is the same as $\M'\, \ind_{(0,\epsilon]}(\Delta_{\rho|\sigma})\,\osig$. Therefore, $\test\kl$ is equivalently defined by the fact that $\L_{\test\kl}$ is the orthogonal projection on $\M'\, \ind_{(0,\epsilon]}(\Delta_{\rho|\sigma})\,\osig$. Since all quantities appearing in this last sentence are well-defined in the standard representation of any von Neumann algebra (see section~\ref{sec:a_primer_in_modular_theory}), this gives a satisfactory starting point for our proof of the extension of Ke Li's lemma.
\medskip

The structure of the paper is as follows. In section~\ref{sec:a_primer_in_modular_theory} we recall the elements of the modular theory of von Neumann algebras required by the statement of our result. In section~\ref{sec:a_general_proof} we give our result and its proof. In section~\ref{sec:discussion} we discuss the merits of our result and its possible applications.

\paragraph{Acknowledgements} YP wishes to thank Nilanjana Datta and Cambyse Rouzé for initiating the pleasant collaboration that led to \cite{DPR} and the present endeavors, and acknowledges the support of Cantab Capital Institute for the Mathematics of Information at the University of Cambridge for a stay during which part of this work was conducted; he also wishes to thank Ke Li and Magdalena Musat for instructive discussions and most of all Yoshiko Ogata who helped chase the modular in Ke Li's original proof. YP was supported by ANR grant ``NonStops'' ANR-17-CE40-0006. SW was also partially supported by the
Fundamental Research Funds for the Central Universities No. FRFCUAUGA5710012222, the NSF of China
No. 12031004, a public Grant as part of the Fondation Mathématique Jacques Hadamard and ANR grant ``ANCG'' ANR-19-CE40-0002.

\section{Modular theory and Haagerup $\Lp p$ spaces} 
\label{sec:a_primer_in_modular_theory}

We recall first a few notions on von Neumann algebras and relative modular operators. Suggested references are \cite{BR1}, and for a pedagogical introduction we recommend section 2 (and in particular sections 2.11 and 2.12) of \cite{JOPP}. We then move on to describe a few properties of Haagerup's $\Lp p$ spaces following \cite{Haagerup} and \cite{Terp}.

Let $\M$ be a von Neumann algebra, i.e.\ a $*$-algebra of bounded linear operators on some Hilbert space, that is closed in the weak operator topology and contains the identity. The set of operators commuting with all elements of $\M$ is called the commutant of $\M$, and denoted $\M'$; von Neumann's bicommutant theorem then states that a $*$-algebra of bounded linear operators containing the identity is a von Neumann algebra if and only if $\M=\M''$, the latter being the bicommutant $(\M')'$ of $\M$.

The notion of nonnegative operators induces a partial order on $\M$: by definition, $X\leq Y$ if $Y-X$ is a nonnegative operator. A map on $\M$ such that the null operator is the only nonnegative operator mapped to the zero element is called \emph{faithful}. Linear forms $\omega$ on $\M$ with the additional regularity property that $\omega(\limsup_n X_n)=\limsup_n \omega(X_n)$ if $X_n$ is an increasing sequence of self-adjoint operators are called \emph{normal}, and this property is also known to be equivalent to various other types of continuity with respect to typical topologies on von Neumann algebras. A linear form which maps nonnegative operators to nonnegative scalars is called \emph{positive}. A positive linear form $\omega$ from nonnegative operators of $\M$ to $[0,+\infty]$ will be called a \emph{weight}. A weight $\omega$ such that the set of nonnegative elements $A$ with $\omega(A)<+\infty$ is weakly dense is called \emph{semifinite}. A weight $\omega$ with the property that $\omega(\id)=1$ is called a \emph{state} (and necessarily takes values in $[0,+\infty)$). A weight or state with the property that $\omega(XY)=\omega(YX)$ for all $X$, $Y$ in $\M$ is called \emph{tracial}, or a \emph{trace}.

A standard representation of some von Neumann algebra $\M$ is a quadruple $(\pi,\H,\H^+,\J)$ where $\H$ is a Hilbert space (with scalar product denoted $\braket\cdot\cdot$), $\pi$ is a faithful morphism of $*$-algebras from $\M$ to $\bh$, $\H^+$ is a self-dual cone of $\H$ (i.e.\ the set of $\phi$ in $\H$ such that $\braket \phi\psi\geq 0$ for all $\psi$ in $\H^+$ is $\H^+$ itself), $\J$ is an anti-unitary involution of $\H$, and these different objects satisfy the following properties:
\begin{myitemize}
  \item $\J\M \J=\M'$,
  \item $\J X\J=X^*$ for $X$ in $\M\cap \M'$,
  \item $\J\psi=\psi$ for $\psi$ in $\H^+$,
  \item $\J X\J X\, \H^+\subset \H^+$ for $X\in \H$.
\end{myitemize}
A standard representation of a von Neumann algebra $\M$ always exists (we will describe one of them below), and if $(\pi_1,\H_1,\H_1^+,\J_1)$ and $(\pi_2,\H_2,\H_2^+,\J_2)$ are two standard representations, then there exists a unitary operator $U:\H_1\to \H_2$ such that $U\pi_1(X)U^*=\pi_2(X)$ for all $X\in \M$, $U\J_1U^*=\J_2$ and $U\H_1^+=\H_2^+$.

We will now describe the modular structure associated with a von Neumann algebra $\M$: fix a standard representation $(\pi,\H,\H_+,\J)$ of $\M$. Then for any normal state $\omega$ on $\M$ there exists a unique $\oome$ in $\H_+$ such that
\begin{equation}\label{eq_defome}
  \omega(X) = \braket{\oome}{\pi(X) \oome} \mbox{ for any }X\in \M.
\end{equation}
This $\oome$ is cyclic in the sense that the closure of $\pi(\M)\H$ is $\H$ itself. If in addition $\omega$ is faithful, then $\oome$ is separating in $\pi(\M)$, that is, $\pi(X)\oome=0$ if and only if $X=0$. We continue by fixing two normal faithful states $\rho$ and $\sigma$. We can then define a densely defined operator $S_{\rho|\sigma}$ by
\[S_{\rho|\sigma} X \osig = X^* \orho.\]
This operator turns out to be closable, and its closure $\overline{S}_{\rho|\sigma}$ has polar decomposition
\[\overline{S}_{\rho|\sigma} = \J {\drs}^{1/2}\]
where $\drs$ is a positive-definite operator on $\H$, which is in general unbounded.

We are now in a position to state our main result, but will first introduce additional elements required for our proof. The reader may wish to take a peek at Theorem~\ref{theo_keli} in section~\ref{sec:a_general_proof}.
\smallskip

To prove the result we consider the Haagerup $\Lp p$-spaces associated with $\M$. We will not need the detailed construction of these spaces and only recall their properties (see chapter 2 of \cite{Terp} for a complete presentation; in the rest of this section all numbered references point to those notes, with e.g.\ Theorem 2.7 meaning Theorem 7 of chapter 2 of \cite{Terp}). An important role will be played by a distinguished von Neumann algebra $\N$, equipped with a normal semifinite faithful trace~$\tau$. This algebra $\N$ then acts on the Hilbert space $\Lp 2(\rr,\H)$ which we simply denote $\K$, for its nature will not matter. For a definition of this algebra $\N$, see the beginning of chapter 2 of \cite{Terp}: between saying too much and saying too little, it seems more convenient to say too little. This $\N$ is such that there exists a faithful normal representation of $\M$ as a sub-von Neumann algebra of~$\N$; in order to spare ourselves an additional notation for this representation, we simply assume that $\M$ is realized as a subalgebra of $\N$, and therefore acts on $\K=\Lp 2(\rr,\H)$. We say that an unbounded operator $K$ on $\K$ is affiliated with $\N$ if any Borel bounded functional calculus of either $\tfrac12(K+K^*)$ or $\tfrac1{2\i}(K-K^*)$ is an element of $\N$.  We say that a densely defined unbounded operator $K$ affiliated with $\N$ is measurable with respect to $(\N,\tau)$ if for every $\delta>0$ there exists a projection $P$ in $\N$ such that the range of $P$ is included in the domain of~$K$, and $\tau(\id -P)<\delta$. Then there exists a subspace $\Lp 1(\M)$ of those measurable operators with respect to $(\N,\tau)$, which is equipped with a faithful positive linear functional $\mathrm{tr}:\Lp 1(\M)\to\cc$ and satisfies the following properties. Among them, the most crucial property that we will use is that $\Lp 1(\M)$ is isometrically isomorphic to a subspace of the tracial weak $\Lp 1$-space on $(\N,\tau)$, and more precisely we have (Lemma~2.5):
\begin{equation}
\tr(X)=\tau\big(\ind_{(1,\infty)}(X)\big) \mbox{ for any } X\in \Lp 1(\M)_{+}, \label{eq:tracial weak norm}
\end{equation} 
where $\Lp 1(\M)_{+}$ denotes the subset of nonnegative operators.

As an illustration, we remark that in the particular case $\M = \mathcal{B}(\mathbb{C}^d )$, the above construction amounts 
$$\Lp 1(\M) =\{x\otimes \xi : x\in \M \} ,\ \N = \mathcal{B}(\mathbb{C}^d ) \otimes \kappa ( \Lp \infty (\mathbb{R}) ), \  \tau = \mathrm{Tr}_d \otimes \kappa (\int \cdot\, e^{-t} \mathrm{d }t ) $$
where $\kappa : \Lp \infty (\mathbb{R}) \to \mathcal{B}(\Lp 2 (\mathbb{R})) $ is the injective homomorphism determined by $\kappa (e ^{\mathrm{i} t \cdot }) f = f(\cdot - t)$ which naturally extends to unbounded $\Lp p$-functions, and $\xi = \kappa ^{-1} ( \exp )$. It is not difficult to show that $\tr (|x\otimes \xi |)= \mathrm{Tr} _d (|x|) $ and hence $x\mapsto x\otimes \xi$ yields an isometric isomorphism between $\Lp 1(\M)$ and the usual noncommutative $\Lp 1$-space on $\mathcal{B}(\mathbb{C}^d ) $ defined via Schatten norms (end of Chapter 2).

For all $1\leq p<\infty$, the space $\Lp p(\M)$ consists of all measurable operators $X$  with respect to $(\N,\tau)$ such that $X=U|X|$ with $U\in\M$ and $|X|^{p}\in \Lp 1(\M)$. Moreover, it holds that
\[
\tr(XY)=\mathrm{tr}(YX) \mbox{ for all } X\in \Lp p(\M),Y\in \Lp q(\M)\mbox{ with }1/p+1/q=1.
\]
If we define $\|X\|_{p}=\tr(|X|^{p})^{1/p}$ then the closed product of measurable operators on these spaces satisfies the Hölder inequality, and in particular we have $\Lp p(\M)\Lp q(\M)\subset \Lp r(\M)$ for any $p,q,r\in[1,+\infty]$ with $1/r=1/p+1/q$. This induces a natural Hilbert space structure on $\Lp 2(\M)$
by taking $p=q=2$. In addition (Theorem 2.7 and Proposition 2.10), $\M=\Lp \infty(\M)$ and the normal states $\rho$, $\sigma$ on $\M$ are of the form 
\begin{equation}\label{eq_matdens2}
  \rho = \tr(\varrho\,\cdot) \qquad \sigma= \tr(\varsigma\,\cdot)
\end{equation}
where $\varrho$, $\varsigma$ are positive elements of $\Lp 1(\M)$ with $\tr(\varrho )  = \tr(\varsigma )=1$. This of course echoes \eqref{eq_matdens} and will allow us to mimic some of the ``spatial'' arguments realized on concrete linear algebras and Euclidean inner products in Ke Li's proof. Taking the values of $p,q,r$ appropriately, we obtain a faithful representation of $\M$ on $\Lp 2(\M)$ as
\[
\pi(X)Y=XY,\quad X\in\M,Y\in \Lp 2(\M).
\]
Denote $\J:X\mapsto X^{*}$ and $\Lp 2(\M)_{+}$ the subset of $\Lp 2(\M)$ made of nonnegative operators. Then $\big(\pi,\Lp 2(\M),\Lp 2(\M)_{+},\J\big)$ is a standard form of $\M$ (see Theorem 2.36). In particular, in this setting $\Omega_\rho = \varrho^{1/2}$ and $\Omega_\sigma = \varsigma^{1/2}$.

Now let $\varrho=\int_{0}^{\infty}\lambda\, \d \erho(\lambda)$ and $\varsigma=\int_{0}^{\infty}\mu\, \d \esig(\mu)$ be spectral decompositions of $\varrho$, $\varsigma$ in $\K$. Then from the definition $S_{\rho|\sigma} X \osig = \mathrm{J}\drs ^{1/2} X \osig = X^* \orho$, we see that 
\[ \drs = \varrho \, \J\varsigma^{-1}\!\J\mbox{ on }\M\osig\]
so that, denoting $\tesig :=\J \esig \J $,
\[\drs = \int_{\rr_+^2} \lambda\mu^{-1}\, \d \erho(\lambda)\, \d \tesig(\mu)\]
at least on $\M\osig$, and since the latter integral expression defines a closed operator, equality holds everywhere. We refer to chapter 5 of \cite{BirmanSolomjak} for the theory of double integrals with respect to commuting spectral measures on Hilbert spaces. In particular, by \cite[Theorem 5.2.6]{BirmanSolomjak} we know that the product of the spectral measures $\erho$ and $\tesig$ is again a spectral measure, and hence the above integral and equality make sense rigorously.


\section{Our result and its proof} 
\label{sec:a_general_proof}
We first state the general form of Ke Li's lemma. As discussed in section~\ref{sec:a_primer_in_modular_theory}, the vector $\osig$ and the operator $\drs$ are meant in any standard representation of $\M$.
\begin{theo}\label{theo_keli}
Let $\M$ be a von Neumann algebra, and $\rho$, $\sigma$ be two faithful normal states on $\M$. For any~$\epsilon>0$ there exists a test $\test\in \M$ such that
  \[\aT\leq \epsilon\qquad \bT\leq \braket{\osig}{\ind_{(\epsilon,+\infty)}(\drs)\osig}.\]
\end{theo}
\begin{rema}
  Our assumption that $\rho$ and $\sigma$ are faithful is unessential, simplifies the proof and the expression of $\drs$ in section~\ref{sec:a_primer_in_modular_theory}, and can be dropped by considering for instance the functional calculus $f(\varsigma)$ with $f(t) = t^{-1}\ind_{t>0}$ instead of $\varsigma^{-1}$.
\end{rema}
The rest of this section is dedicated to the proof of Theorem~\ref{theo_keli}. We consider the standard representation $\big(\pi,\Lp 2(\M),\Lp 2(\M)_{+},\J\big)$ of $\M$ that we discussed in the second part of section~\ref{sec:a_primer_in_modular_theory}. Once again we simply write $\M$ for $\pi(\M)$, so that $\M$ will act by left-multiplication on $\Lp 2(\M)$.

Let $\pp$ be the following orthogonal projection operator on $\Lp 2(\M)$:
\begin{equation*}
  \pp = \ind_{(0,\epsilon]}(\drs).
\end{equation*}
We recall that if $X$ is an element of $\Lp2(\M)$ then its \emph{support} in $\M$, which we denote by $\ell(X)$ (with an $\ell$ for ``left support''), is defined as the smallest orthogonal projector $\proj$ in $\M$ such that $\proj X=X$. Equivalently, it is the projection onto the closure of the range of $X$ (see the explanations around \cite[Chapter 5, Definition 1.4]{Takesaki}), which is exactly $\overline{X\M}$ in our setting. It is immediate to see that this projector has range $\overline{\M'X}$ since $PX\M =X\M = \M ' X $ must be included in the range of $P$. Similarly we define the support of $X \in \Lp2(\M)$ in $\M'$, which we denote by $r(X)$ (with an $r$ for ``right support'')  as the smallest orthogonal projector $\proj'$ in $\M'$ such that $\proj' X=X$, which may be equivalently defined to be the projection onto the range of $X^*$. \label{page_defell}

First define our test $\test$ as the \emph{support} in $\M$ of $\pp(\osig)$. By definition $\pp(\osig)$ is an element of $\Lp 2(\M)$, and therefore a densely defined closed operator on $\K$ which is affiliated with $\N$; the polar decomposition of $\pp(\osig)$ as an operator on $\K$ has partial isometry $U$ which from the properties of $\Lp 2(\M)$ belongs to $\M$. Then $UU^*$ is by definition the smallest orthogonal projector $\proj$ in $\cB(\K)$ such that $\proj\,\pp(\osig)=\pp(\osig)$, and belongs to $\M$. Therefore $T$ is equivalently defined as the orthogonal projector onto $\overline{\M'\pp(\osig)}$, as $UU^*$, or as left-multiplication by the orthogonal projector onto the closed range of $\pp(\osig)$ when the latter is viewed as an operator on $\K$. Note that by definition, $\test\in \M$.

From the first expression for $\test$ we have
\[\sigma(\id - \test) = \norme{\osig-\test \osig}^2 \leq \norme{\osig-\pp(\osig)}^2=\braket{\osig}{\ind_{(\epsilon,+\infty)}(\drs) \osig}\]
where:
\begin{myitemize}
  \item the first equality follows from \eqref{eq_defome} and the fact that $\id-\test$ is a self-adjoint projection,
  \item the inequality follows from the inequality $\norme{\Psi - \test \Psi}\leq \norme{\Psi - \Phi}$, valid for any $\Psi$ in~$\Lp 2(\M)$ and $\Phi\in \ran \test$ by definition of an orthogonal projection, and the fact that $\pp(\osig)$ is an element of $\M'\pp(\osig)\subset\ran\test$,
  \item the second equality follows from \eqref{eq_defome} and the fact that $\id-\pp$ is a self-adjoint projection.
\end{myitemize}
This proves the second bound in Theorem~\ref{theo_keli}, and we now move on to prove the first bound.
\smallskip

Recall the equality in the previous section 
\[\drs = \int_{\rr_+^2} \lambda\mu^{-1}\, \d \erho(\lambda)\, \d \tesig(\mu).\]
Approximating $\mathbf{1}_{(0,\epsilon]}$ by polynomials weakly, it is easy to see that
\[
\pp=\int_{\rr_+^2} \ind_{(0,\epsilon]}(\lambda\mu^{-1})\, \d \erho(\lambda) \d \tesig(\mu).
\]
Therefore
\begin{align}
\pp(\osig)
  &=\int_{\rr_+^2} \ind_{(0,\epsilon]}(\lambda\mu^{-1})\mu^{1/2}\, \d \erho(\lambda) \d \esig(\mu)\nonumber\\
  &= \int_0^\infty \mu^{1/2} \int_0^\infty  \ind_{(0,\epsilon \mu ]}(\lambda) \d \erho(\lambda) \d \esig(\mu)\nonumber   \\
  &=\int_{0}^{\infty}\mu^{1/2}\ind_{(0,\epsilon\mu]}(\varrho)\, \d \esig(\mu). \label{eq_exptest}
\end{align}
We remark that the integral with respect to a spectral measure is by definition understood as approximations by Riemann sums  (see e.g. \cite[IX.1.10 and X.4.6]{Conway}), and in particular all above integrals are understood and verified in this way.
\smallskip

We now prove an easy regularity lemma that will allow us to approximate $T$ thanks to Riemann sums approximating the expression \eqref{eq_exptest} for $\pp (\osig)$. 
\begin{lemm}
	Let $X\in \Lp 2 (\M)$ and assume that $(X_n)_n$ is a sequence of bounded operators on $\K$ that converges to $X$ strongly. Let $\sproj_n$ (respectively $\sproj$) be the support of $X_n$ (respectively $X$) as an operator on $\K$. Denote by $\proj_n$ (respectively $\proj$) the operator of left-multiplication by $\sproj_n$ (respectively $\sproj$) on $\Lp 2 (\M)$. Then
	\begin{equation}
		\rho (P)  \leq \limsup_n \rho (P_n) \label{eq_regularite}.
	\end{equation} 
\end{lemm}

\begin{proof}
	For any $\xi$ in $\K$ we have using $ \sproj_n \, X_n \xi =  \, X_n \xi$ and $\sproj\, X\xi=X\xi$:
\begin{align*}
\| \sproj_n \sproj X\xi-\sproj X\xi\|\leq\norme{\sproj_n  X\xi - \sproj_n  X_n\xi}+\norme{\sproj_n  X_n\xi -   X\xi}\leq 2\norme{X_n\xi - X\xi}
\end{align*}
and the upper bound goes to zero as $n\to\infty$. Recall that $p$ is the projection onto the closure of the range of $X$, so $p\eta =0$ for $\eta\in (X \mathcal K )^\bot$. In particular, the above inequality and limit yield that $ \sproj_n\sproj$ converges to $\sproj$ strongly. Consequently $p p_n p$ also converges to $p$ strongly. Note that $\langle   (p_n p - p) \xi,\eta\rangle = \langle    \xi, (p p_n - p) \eta\rangle$, so  $p p_n$ converges to $p$ weakly. Altogether these imply that $(1-p) p_n p$ and $p p_n (1-p)$ converge to $0$ weakly. Now consider $\proj_n$ (respectively $\proj$) the operator of left-multiplication by $\sproj_n$ (respectively $\sproj$) on $\Lp 2 (\M)$. Note that the normality of $\rho$ is equivalent to the ultraweak continuity on $\M$, or equivalently the weak continuity on the unit ball of $\M$, and note also that the families $\{PP_n P\}, \{(1-P) P_n P\}, \{P P_n (1-P)\}$ are all bounded. Hence we get
\begin{align*}
\rho (P)& = \lim_n \rho (P P_n P) \nonumber\\
& = \lim_n \rho (P P_n P) + \lim_n \rho ((1-P) P_n P) + \lim_n \rho (P P_n (1-P)) \nonumber\\
&= \lim_n \big(\rho (P_n  ) - \rho((1-P) P_n (1-P)) \big) \nonumber\\
&\leq \limsup_n \rho (P_n).
\end{align*}  
\end{proof}

Recall now that our test $\test$ is defined as the support $\ell(X)$ of $X:=\pp (\osig)$. As explained after \eqref{eq_exptest}, we may choose a sequence of Riemann sums $(X_n)_n$ of the form 
\[X_n=\sum_{i}\mu_{i}^{1/2}\ind_{[0,\epsilon\mu_{i})}(\varrho)\,\big(\esig(\mu_i)-\esig(\mu_{i-1})\big)
\]
that converges strongly to $X$, where $\esig(\mu):= \esig((-\infty,\mu])$ denotes the spectral measure of the interval $(-\infty,\mu]$. Let $\test_n$ be the support $\ell(X_n)$ of $X_n$. From \eqref{eq_regularite}, we have $\rho(\test)\leq \limsup_n \rho(\test_n)$ and we therefore consider $\rho(\test_n)$. Note that $T_n$ is the projection onto the closure of range of $X_n$, which, by the definition of $X_n$, is contained in the span of the ranges of $\ind_{[0,\epsilon\mu_{i})}(\varrho)\,\big(\esig(\mu_i)-\esig(\mu_{i-1})\big)$. Hence
\[
T_n\leq\sum_{i}\ell\big(\ind_{[0,\epsilon\mu_{i})}(\varrho)\,\big(\esig(\mu_i)-\esig(\mu_{i-1})\big)\big)=: \test_n',
\]
where we omitted the scalar coefficients $\mu_{i}^{1/2}$ since $X$ and its multiple $\lambda X$ must have the same range for any $X\in \Lp 2(\M)$. Note that $\ind_{[0,\epsilon\mu_{i})}(\varrho)$ is a projection whose range contains that of $\ell\big(\ind_{[0,\epsilon\mu_{i})}(\varrho)\,\big(\esig(\mu_i)-\esig(\mu_{i-1})\big)\big)$, so we write 
\[
\ell\big(\ind_{[0,\epsilon\mu_{i})}(\varrho)\,\big(\esig(\mu_i)-\esig(\mu_{i-1})\big)\big)=\ind_{[0,\epsilon\mu_{i})}(\varrho)\,\ell\big(\ind_{[0,\epsilon\mu_{i})}(\varrho)\,\big(\esig(\mu_i)-\esig(\mu_{i-1})\big)\big).
\]
Note that (see e.g.\ \cite[2.5 and 2.9]{FackKosaki})
\begin{equation*} \label{eq_monotonietauind}
\tau\big(\ind_{(1,\infty)}(X)\big)\leq\tau\big(\ind_{(1,\infty)}(Y)\big) \ \mbox{ if }\ 0\leq X\leq Y.
\end{equation*}
Thus together with \eqref{eq:tracial weak norm},
\begin{align*}
\tr(\varrho \test_n') & =\sum_{i}\tr\big(\varrho \ind_{[0,\epsilon\mu_{i})}(\varrho)\,\ell\big(\ind_{[0,\epsilon\mu_{i})}(\varrho)\,\big(\esig(\mu_i)-\esig(\mu_{i-1})\big)\big)\big)\\
 & =\sum_{i}\tau\Big(\ind_{(1,\infty)}\Big(\varrho \ind_{[0,\epsilon\mu_{i})}(\varrho)\,\ell\big(\ind_{[0,\epsilon\mu_{i})}(\varrho)\,\big(\esig(\mu_i)-\esig(\mu_{i-1})\big)\big)\Big) \Big)\\
 & \leq\sum_{i}\tau\Big(\ind_{(1,\infty)}\Big(\epsilon\mu_{i}\,\ell\big(\ind_{[0,\epsilon\mu_{i})}(\varrho)\,\big(\esig(\mu_i)-\esig(\mu_{i-1})\big)\big)\Big) \Big)\\
 & \leq\sum_{i} \ind_{(1,\infty)}(\epsilon\mu_{i})\, \tau\Big(\ell\big(\ind_{[0,\epsilon\mu_{i})}(\varrho)\,\big(\esig(\mu_i)-\esig(\mu_{i-1})\big)\big)\Big).
\end{align*}
Note that $\sproj_{i,1}:=\ind_{[0,\epsilon\mu_{i})}(\varrho)$ and $\sproj_{i,2}:=\esig(\mu_i)-\esig(\mu_{i-1})$ are projections\footnote{By abuse of notation, here $p_{i,2}$ is understood as an element in $\mathcal M \subset \mathcal N$ by right multiplication action on the unit of $\mathcal M$}, and that $\ell(\sproj_{i,1}\,\sproj_{i,2})$ is equivalent to $r(\sproj_{i,1}\,\sproj_{i,2})$ (see e.g. \cite[Chapter V, Proposition 1.5]{Takesaki}), i.e.\ that there exists a unitary $U$ in $\M$ such that $\ell(\sproj_{i,1}\,\sproj_{i,2})=UU^*$ and $r(\sproj_{i,1}\,\sproj_{i,2})=U^*U$. But  $r(\sproj_{i,1}\,\sproj_{i,2})$ is by definition the projection onto the range of $\sproj_{i,2}\,\sproj_{i,1}$ which is contained in that of $\sproj_{i,2}$.
Thus, $\ell(\sproj_{i,1}\,\sproj_{i,2})$ is equivalent to a subprojection of $\sproj_{i,2}$, whence $\tau\big(\ell(\sproj_{i,1}\,\sproj_{i,2})\big)\leq\tau(\sproj_{i,2})$. So the previous inequality reads
\begin{align*}
\tr(\varrho \test_n') & \leq \tau\Big(\sum_{i} \ind_{(1,\infty)}(\epsilon\mu_{i})\, (\esig(\mu_i)-\esig(\mu_{i-1}))\Big).
\end{align*}
Taking the limit of Riemann sums yields the integral with respect to the spectral measure, $\int\ind_{(1,\infty)}(\epsilon\mu)\, \d \esig(\mu)$, which equals the functional calculus $\ind_{(1,\infty)}(\epsilon\varsigma)$ (see e.g. \cite[X.4.11]{Conway}). Using \eqref{eq:tracial weak norm} again, we get
\[
\tr(\varrho \test)\leq\tau\Big(\int\ind_{(1,\infty)}(\epsilon\mu)\, \d \esig(\mu)\Big)=\tau\big(\ind_{(1,\infty)}(\epsilon\varsigma)\big)=\tr(\epsilon\varsigma)=\epsilon.
\]
The proof is complete.\qed

\section{Applications and comparison of the Ke Li and Neyman--Pearson tests} 
\label{sec:discussion}

In most of this section we return to the practical task of interest, which is of discriminating between the states $\rho$ and $\sigma$. As we mentioned in the introduction, in such discrimation tasks one will typically try to make both error probabilities $\aT$ and $\bT$ small (recall that $\aT$ and $\bT$ are defined by \eqref{eq_defalphabeta}). However, it is expected that there is a tradeoff between one error probability and the other. One must therefore make more precise in what sense we want to ``make both $\aT$ and $\bT$ small''. One possible sense is natural if we assume that prior probabilities $p$ and $q:=1-p$ can be assigned to the states $\rho$ and~$\sigma$ respectively. One may then decide to minimize the quantity $p\, \alpha(T)+q\, \beta(T)$. This is the realm of symmetric hypothesis testing, as opposed to asymetric hypothesis testing where one wishes e.g.\ to minimize $\aT$ under a specific constraint on $\bT$.
\smallskip


To first state the relevant result for symmetric hypothesis testing we define a specific test $\test\np$ called the \emph{Neyman--Pearson} test. In the finite-dimensional case this test is defined as (recall that the left support $\ell$ was defined on page \pageref{page_defell})
\[\test\np := \ell\big(\ind_{\rr_+}(p\varrho-q\varsigma)\big).\]
In the general case the definition is formally similar:
\[\test\np := \ell\big((p\rho-q\sigma)_+\big)\]
but here one needs to rely on the (unique) Jordan decomposition
\[p\rho-q\sigma= (p\rho-q\sigma)_+ - (p\rho-q\sigma)_-\]
with $(p\rho-q\sigma)_\pm$ two positive normal forms on $\M$. We will freely denote $\test\np$ by $\test\np(p,q)$ when we need to emphasize the dependency on $p,q$.

The following result is known as the quantum Neyman--Pearson lemma and was proved in \cite{ANS} in the finite-dimensional case and in \cite{JOPS} in the general case:
\begin{equation} \label{eq_np}
	p\, \alpha(\test\np)+q\, \beta(\test\np)= \inf_T\big( p\, \alpha(T)+q\, \beta(T)\big)
\end{equation}
where the infimum is indifferently over the set of orthogonal projections of $\M$, or over the set of elements $T$ of $\M$ satisfying $0\leq T \leq \id$. The associated relevant upper bound was also proven in \cite{ANS,JOPS} and is called the \emph{Chernoff bound}:
\begin{equation} \label{eq_chernoff}
	p\, \alpha(\test\np)+q\, \beta(\test\np)\leq \inf_{s\in[0,1]}\, p^sq^{1-s}\, \braket{\osig}{\drs^s \osig}.
\end{equation}
\smallskip

Denote by $\test\kl$ (or, again, by $\test\kl(\epsilon)$ when we need to emphasize the dependence on~$\epsilon$) the test constructed in section~\ref{sec:a_general_proof} (we confess a slight inconsistency in our choice of notation, as the present $\test\kl$ is the $\L_{\test\kl}$ of section~\ref{sec:introduction}), which we call the Ke Li test. A first interesting observation is that, in the case where $\M$ is commutative, 
then the Neyman--Pearson and Ke Li test coincide in the sense that 
\begin{equation} \label{eq_tkltnp}
	\test\kl(\epsilon) = \test\np\big(\frac1{1+\epsilon},\frac\epsilon{1+\epsilon}\big).
\end{equation}
This can be easily seen in the finite-dimensional case, i.e.\ when $\M=\{L_X, \, X\in \cB(\cc^n)\}$ as described in section~\ref{sec:introduction} (and using the same notation). Using the fact that the families $(a_x)_x$ and $(b_y)_y$ are the same (up to permutation) when $\varrho$ and $\varsigma$ commute, expression \eqref{eq_pomegarho} takes the form
\begin{equation} \label{eq_comparaison1}
	\ind_{(\epsilon,+\infty)}(\Delta_{\rho|\sigma})\,\osig=\sum_{x=1}^n \ind_{\lambda_x> \epsilon \mu_x}\, \mu_x^{1/2} \ketbra{a_x}{a_x}.
\end{equation}
On the other hand, when $p =\frac1{1+\epsilon}$,
\begin{equation} \label{eq_comparaison2}
	\ind_{\rr_+}(p\varrho-q\varsigma)=\sum_{x=1}^n \ind_{\lambda_x> \epsilon \mu_x}\, \ketbra{a_x}{a_x}.
\end{equation}
Therefore, the supports of the operators in \eqref{eq_comparaison1} and \eqref{eq_comparaison2} are the same, which yields identity~\eqref{eq_tkltnp}.
\medskip


Now let us compare the merits of Theorem~\ref{theo_keli}, in comparison to the Chernoff bound \eqref{eq_chernoff}. From now on we always consider $p =\frac1{1+\epsilon}$ so that $p$ satisfies the relation $q/p=\epsilon$. The Chernoff bound \eqref{eq_chernoff} implies
\begin{equation}\label{eq_majnp}
	\rho(\test\np)\leq \epsilon \qquad
	\sigma(\test\np)\leq \epsilon^{-s} \braket{\osig}{\drs^s \osig}
\end{equation}
whereas Theorem~\ref{theo_keli} gives
\begin{equation}\label{eq_majkl}
	\rho(\test\kl)\leq \epsilon\qquad
	\sigma(\test\kl)\leq \braket{\osig}{\ind_{(\epsilon,+\infty)}(\drs)\osig}.
\end{equation}
If the upper bound for $\sigma(\test\kl)$ in \eqref{eq_majkl} is bounded by an application of Markov's exponential inequality, then \eqref{eq_majnp} and \eqref{eq_majkl} yield the same pair of bounds. It therefore turns out that the estimates \eqref{eq_majkl} given by Theorem~\ref{theo_keli} are no better than those \eqref{eq_majnp} given by the Chernoff bound, unless one has an estimate for $\braket{\osig}{\ind_{(\epsilon,+\infty)}(\drs)\osig}$ more precise than that given by Markov's exponential inequality. It therefore seems that our result is better suited to situations where specific information on the tails of the distribution of $\drs$ with respect to the state $\osig$ is available. To describe why we expect this to be the case in most situations, let us describe the asymptotic setting typical of Stein's lemma.

Consider two sequences $(\rho_n)_n$ and $(\sigma_n)_n$ of states on possibly different algebras~$\M_n$, which one would like to discriminate ``for large $n$''. One will therefore try and construct a sequence $(\test_n)_n$ of tests $\test_n\in\M_n$ for which $\big(\alpha(\test_n)\big)_n$ and $\big(\beta(\test_n)\big)_n$ are ``small'' in some sense. Stein's lemma aims at answering the following question: if one only insists on having $\sup_n\beta(\test_n)\leq \epsilon$, what is the best rate of decrease of $\alpha(\test_n)$ one can hope for? The standard result is that under some conditions, an exponential decrease of any rate $r$ strictly smaller than a quantity $D$ called the \emph{relative entropy} of $(\rho_n)_n$ and $(\sigma_n)_n$ is possible (the ``existence'' statement), but an exponential decrease of a rate strictly larger than $D$ is not possible (the ``optimality'' part). It is well-known that the Chernoff bound \eqref{eq_chernoff} is enough to prove a non-trivial version of the existence part of Stein's lemma (see \cite{JOPS,NussbaumSzkola}). We now show that Theorem~\ref{theo_keli} is enough to prove the existence part of Stein's lemma. Let us therefore, for every $n$ consider the Ke Li test $T_n$ associated with
\[\epsilon=\exp \big(-D(\rho_n|\sigma_n)+ w_n \delta\big)\]
where $w_n \to\infty$ is a ``growth scale'' associated with the model, $\delta$ is some positive number, and we define
\[D(\rho_n|\sigma_n) := \braket{\Omega_{\sigma_n}}{\log \Delta_{\rho_n|\sigma_n} \, \Omega_{\sigma_n}}.\]
We immediately have
\begin{equation}
	\log \alpha(T_n)\leq -D(\rho_n|\sigma_n) + w_n \delta\
\end{equation}
and
\begin{align}
	\beta(T_n) &= \braket{\Omega_{\sigma_n}}{\ind_{(-\infty, \,D(\rho_n|\sigma_n) - w_n \delta)} (\log \Delta_{\sigma_n|\rho_n})\,\Omega_{\sigma_n}} \notag\\
	&= \braket{\Omega_{\sigma_n}}{\ind_{(-\infty, -\delta)} \Big(\frac{\log \Delta_{\sigma_n|\rho_n} - D(\rho_n|\sigma_n)}{w_n}\Big)\,\Omega_{\sigma_n}} \label{eq_preuvestein}.
\end{align}
Denote now for all $n$ by $X_n$ a random variable on some probability space $(\Omega,\mathbb P)$ with the same distribution as $\log \Delta_{\sigma_n|\rho_n}$ in the state $\Omega_{\sigma_n}$, i.e.\ a random variable satisfying
\[\mathbb E\big(\varphi(X_n)\big) = \braket{\Omega_{\sigma_n}}{\varphi(\log \Delta_{\sigma_n|\rho_n})\,\Omega_{\sigma_n}}\]
for any smooth bounded function $\varphi$. This $X_n$ is known in the quantum information community as a Nussbaum-Szko\l a random variable (see \cite{NussbaumSzkola}). It now follows from~\eqref{eq_preuvestein} the existence part of the quantum Stein lemma under milder conditions than in \cite{JOPS} where the existence of a pressure function (i.e.\ existence of a growth rate for exponential moments) is assumed. 
\begin{theo}
	If $D\coloneqq \lim_n \frac 1 {w_n} D(\rho_n|\sigma_n) $ exists and if the sequence $(X_n)_n$ satisfies a weak law of large number $\frac {X_n - D(\rho_n|\sigma_n)}{w_n} \to 0$ (convergence in probability), then
\[\limsup_n \frac 1 {w_n}\log \alpha(T_n)\leq -D + \delta\quad\mbox{and}\quad \beta(T_n) \to 0,\]
\end{theo}
Note in particular that in the ``i.i.d.\ case'' where for each $n$ one has 
\[ \M_n = \mathcal B(\cc)^{\otimes n}\qquad \rho_n=\rho^{\otimes n}\qquad \sigma_n=\sigma^{\otimes n}\]
in which case $D(\rho_n|\sigma_n)= n \,D(\rho|\sigma)$ and $X_n$ is the sum of $n$ identical, identically distributed bounded random variables, so that the above assumptions hold for the growth function $w_n=n$.
\medskip

It is clear, on the other hand, that under stronger assumptions on the asymptotic behavior of the sequence $(X_n)_n$, typicall in the form of a central limit theorem, the same strategy can give more information on the behaviour of $\beta(T_n)$. Our proof above shows that under the sole condition that $\beta(T_n) \to 0$, one can have $\alpha(T_n)$ decrease exponentially with any rate strictly smaller than $D$. The papers \cite{KeLi,DPR,DR} ask whether a rate \emph{equal} to~$D$ is possible, and what is then the ``next order'' describing the optimal rate of decrease for $\beta(\test_n)$. In particular, these papers identify conditions under which one can construct a sequence $(T_n)_n$ satisfying both conditions $\sup_n\alpha(\test_n)\leq \epsilon$ and 
\begin{equation}\label{eq_secondorder}
	-\log\beta(\test_n) = nD + \sqrt n\, t+ o(\sqrt n)
\end{equation}
for any $t < \Phi^{-1}(\epsilon) V $. Here $\Phi$ is the cumulative distribution function of a standard normal distribution, $V$ is a quantity depending on the sequences $(\rho_n)_n$ and $(\sigma_n)_n$. This is what we call the ``existence'' part of a second-order Stein's lemma.

The articles \cite{KeLi,DPR,DR} use additional information on the sequences $(\rho_n)_n$ and $(\sigma_n)_n$ to gain more information on the behaviour of $(X_n)_n$ and (in the case where every $\M_n$ is a finite-dimensional algebra) prove the second order behaviour \eqref{eq_secondorder}. They also prove the ``optimality'' result that if \eqref{eq_secondorder} with $t < \Phi^{-1}(\epsilon) V$ then $\alpha(\test_n)\to 1$, to give a full second-order Stein's lemma.

For instance, the article \cite{DR} used a submultiplicativity property of the sequences $(\rho_n)_n$ and $(\sigma_n)_n$ to derive concentration inequalities for the distribution of $(X_n)_n$. The nature of the considered sequences $(\rho_n)_n$ and $(\sigma_n)_n$, however, imposes the assumption that every $\M_n$ is finite-dimensional; in that case Theorem~\ref{theo_keli} follows from Ke Li's original result and the present general extension is not needed. 

The articles \cite{KeLi,DPR} also prove a full second-order Stein's lemma either in the ``i.i.d.\ case'' (in \cite{KeLi}) or under a more general assumption on the exponential moments\footnote{This assumption is mathematically an overkill but the behaviour of exponential moments is typically considered in practical studies of models from statistical mechanics} of $(X_n)_n$ that yield a central limit theorem for this sequence. The proof in \cite{DPR}, however, only relies on Theorem~\ref{theo_keli} and, for the optimality part, on a ``lower Chernoff bound'' which was proved in the general case in \cite{JOPS}. Therefore, the present proof of Theorem~\ref{theo_keli} immediately extends Theorem 1 of \cite{DPR} to the case of general von Neumann algebras, showing a full second-order Stein's lemma under assumptions formally similar to those in \cite{DPR}. More precisely, for the two aforementioned sequences of states $(\rho_n)_n, (\sigma_n)_n$, we may also consider the quantum information variance $V(\rho_n  | \sigma_n )=Var (X_n )$ and the quantity $\Psi_s (\rho_n | \sigma_n ) = \log \mathbb{E} (e^{-sX_n })$ for $s\in [0,1]$, and introduce verbatim the natural Condition 1 of \cite{DPR} with these general definitions in our new context; we omit the lengthy technical expressions of this condition and refer the interesting readers to the original text \cite{DPR}.  We assume that $ (\rho_n)_n$ and $  (\sigma_n)_n$ satisfy this condition with respect to the scalars $(w_n)_n$, and 
\[D\coloneqq \lim_n \frac 1 {w_n} D(\rho_n|\sigma_n) ,\quad V \coloneqq \lim_{n\to\infty } \sqrt{\frac{1}{w_n} V(\rho_n \| \sigma_n )}\]
exist. 
We may then state the following result on second order asymptotics.
\begin{theo}
	Fix $\varepsilon\in (0,1)$ and $\beta_n (\varepsilon) \coloneqq \inf_{0\leq T_n \leq 1 } \{\beta(T_n) : \alpha (T_n)\leq \varepsilon \}$. Then for any $t >  \Phi ^{-1} (\varepsilon) V $, 
	\[-\log \beta_n (\varepsilon) \leq  D + w_n t + o (w_n ), \]
	and for any $t_2 < \Phi ^{-1} (\varepsilon) V $, 
	\[-\log \beta_n (\varepsilon) \geq  D + w_n t  + o (w_n ). \]
\end{theo}
Following verbatim the arguments in \cite[Section 4.1]{DPR}, in the particular i.i.d. case $\rho_n= \rho^{\otimes n}$ and $\sigma_n= \sigma^{\otimes n}$ we obtain:
\begin{theo}
	Fix $\varepsilon\in (0,1)$ and $\beta_n (\varepsilon)$ as above, and let $\rho_n= \rho^{\otimes n}$ and $\sigma_n= \sigma^{\otimes n}$ be the states on von Neumann algebras $\M^{\otimes n}$ for two normal states $\rho$ and $\sigma$. Then 
	\[-\log \beta_n (\varepsilon) = n D(\rho  | \sigma ) +  \sqrt{n V(\rho  | \sigma )  } \Phi ^{-1} (\varepsilon) + O (\log n ). \]
\end{theo}

It is expected that such extensions of second-order Stein's lemma can also provide similar extensions for other information theoretic tasks which are typically reduced to hypothesis testing, such as coding problems, but we will not discuss the detailed extensions at this time. One can also remark that, in the same way that (the first-order) Stein's lemma gave the mathematical quantity known as entropy production a physical meaning as the optimal rate for determination of the arrow of time (see Section 7.4 of \cite{JOPS}), our second-order Stein's lemma gives a physical meaning to a new quantity deriving from the $V$ mentioned above. We leave the investigation of this fact to the interested reader. 

\end{document}